\documentclass[12pt,sort&compress]{article}

\usepackage[T1]{fontenc}
\usepackage[latin9]{inputenc}
\usepackage{amsmath}
\usepackage{amssymb}
\usepackage{graphicx}
\usepackage[numbers]{natbib}

\makeatletter
\newcommand{\lyxaddress}[1]{
\par {\raggedright #1
\vspace{1.4em}
\noindent\par}
}

\@ifundefined{date}{}{\date{}}

\usepackage{indentfirst}
\usepackage{amsfonts}
\usepackage[T1]{fontenc}
\usepackage{ae,aecompl}
\usepackage{sidecap}
\usepackage[section]{placeins}
\usepackage{epsf}

\makeatother

\begin{document}

\title{Calculation of the $\gamma/\gamma^{\prime}$ interface free energy
in the Ni-Al system by the capillary fluctuation method}

\author{Y. Mishin%
\thanks{Email: ymishin@gmu.edu%
}}

\maketitle

\lyxaddress{School of Physics, Astronomy and Computational Sciences, MSN 3F3,
George Mason University, Fairfax, Virginia 22030, USA}
\begin{abstract}
Monte Carlo computer simulations with an embedded-atom potential are
applied to study coherent $\gamma/\gamma^{\prime}$ interfaces in
the Ni-Al system. The (100) interface free energy has been extracted
from the power spectrum of equilibrium shape fluctuations (capillary
waves) and found to decrease with temperature from about 20 mJ/m$^{2}$
at 550 K to about 10 mJ/m$^{2}$ at 1200 K. These numbers are in reasonable
agreement with existing experimental data. Strengths and disadvantages
of the capillary wave method are discussed. 
\end{abstract}

\bigskip{}

Interfaces between different phases control microstructural stability
of many precipitation-hardened alloys. In particular, the coarsening
kinetics of $\gamma^{\prime}$-phase precipitates in the $\gamma$-phase
matrix in Ni-based superalloys is largely controlled by the free energy
$\sigma$ of the $\gamma/\gamma^{\prime}$ interfaces \citep{Sims87}.
Because reliable computational predictions of $\sigma$ for multi-component
commercial alloys are extremely difficult, attention has been focused
on $\gamma/\gamma^{\prime}$ interfaces in the binary Ni-Al system
as a model of superalloys. In the latter case, the $\gamma$-phase
is an atomically disordered Ni-based solid solution whereas the $\gamma^{\prime}$-phase
is the intermetallic compound Ni$_{3}$Al (L1$_{2}$ structure) accommodating
a few atomic per cent of off-stoichiometry. Even in this simpler case,
however, calculations $\sigma$ are challenging due to the complexity
of the underlying interface thermodynamics \citep{Frolov2012a,Frolov2012b},
issues with reliability of atomic interaction models, and the sheer
volume of computations \citep{Mishin04a,Mishin2010a,Frolov2012a,Frolov2012b}.

To further simplify the problem, several authors focused on coherent
interfaces between pure Ni and perfectly stoichiometric Ni$_{3}$Al
employing either first-principles calculations \citep{Wang2009,Mao2012a}
or atomistic simulations with empirical potentials \citep{Yashiro2002,Mishin04a,Zhu2005}.
In particular, Mao et al.~\citep{Mao2012a} applied density functional
theory calculations to compute $\sigma$ as a function of temperature
for three different low-index orientations of the interface. Their
calculations include the ferromagnetic ordering, coherency strain
energy and atomic vibrations. They predict that $\sigma$ significantly
decreases with temperature. For example, the (100) interface free
energy decreases from 27 mJ/m$^{2}$ at 0 K to about 23 mJ/m$^{2}$
at the temperature of 1000 K, demonstrating that the temperature effect
should be taken into account for accurate calculations of $\sigma$.

An equally important factor that should also be taken into account
is that at finite temperatures, the $\gamma$-phase is not pure Ni
and the $\gamma^{\prime}$-phase is not stoichiometric Ni$_{3}$Al.
The actual chemical compositions of both phases and long-range order
in the$\gamma^{\prime}$-phase vary with temperature according to
the Ni-Al phase diagram \citep{Massalski}. First-principles-based
calculations of thermodynamic equilibrium between the $\gamma$ and
$\gamma^{\prime}$ phases could be pursued by the cluster expansion
approach \citep{Sluiter98,Asta01,vandewalle09}. An alternate, and
computationally more efficient, approach is offered by atomistic simulations
with empirical potentials \citep{Mishin04a,Mishin2010a}. For coherent
$\gamma/\gamma^{\prime}$ interfaces, accurate but rather tedious
calculations of $\sigma$ could be performed by the thermodynamic
integration method based on rigorous interface thermodynamic relations
developed in \citep{Frolov2012a,Frolov2012b}. In this paper we explore
a different route involving the analysis of capillary waves on the
interfaces.

For rough interfaces supporting capillary waves, their equilibrium
power spectrum can be analyzed to extract the interface free energy,
or more accurately, the interface stiffness $(\sigma+\sigma^{\prime\prime})$,
where $\sigma^{\prime\prime}$ is the second derivative of $\sigma$
with respect to the interface inclination angle. For a narrow, ribbon-like
interface geometry, the mean-squared amplitude (power) of capillary
waves at a given temperature $T$ is given by \citep{Hoyt01} 
\begin{equation}
\left\langle a^{2}(k)\right\rangle =\dfrac{k_{B}T}{(\sigma+\sigma^{\prime\prime})Ak^{2}},\label{eq:1}
\end{equation}
where $k_{B}$ is Boltzmann's factor, $k$ is the wavenumber of the
wave, and $A=wL$ is the interface area with $L$ being the interface
length and $w\ll L$ the interface ribbon width. Using the computed
power spectrum, $(\sigma+\sigma^{\prime\prime})$ can be determined
from the slope of the plot $1/\left\langle a^{2}(k)\right\rangle $
versus $k^{2}$. The capillary fluctuation method was applied to compute
the free energies and stiffnesses of solid-liquid interfaces \citep{Hoyt01,Morris02a,Morris03,Hoyt03a,Davidchack06,Becker07a,Amini08,Becker09,Rozas2011}
and grain boundaries in single-component systems \citep{traut05,Foiles06a,Karma-2012a}. 

In this paper we apply this method to solid-solid phase boundaries
in a binary system. This extension required solving the following
problem. The previous applications of the method employed molecular
dynamics simulations to measure the equilibrium fluctuation spectrum.
Due to the well-known computational limitations, molecular dynamics
simulations are usually run for times no longer than tens of nanoseconds.
This time scale is too short to observe a significant amount of substitutional
solute diffusion in the crystalline lattice even at high temperatures
approaching the melting point. As a consequence, the computed capillary
wave spectrum does not include fluctuation modes requiring diffusion-controlled
redistribution of the solute between the crests and troughs of the
waves. A possible answer would be to resort to lattice-free semi-grand
canonical Monte Carlo simulations \citep{FrenkelS02}. In this case,
redistribution of the solute occurs by an artificially fast process
involving random changes of the chemical species of the atoms. Although
the dynamics of this process are unphysical, the simulation quickly
and correctly samples numerous configurations representing the equilibrium
state of the systems. This permits a fast calculation of the ensemble
average value of $\left\langle a^{2}(k)\right\rangle $. In practice,
however, it is impossible to adjust the chemical-potential difference
$\Delta\mu$ between the species to \emph{exactly} match the phase
equilibrium condition. Consequently, the interface will always migrate
towards one of the phases, creating a dynamic situation that may affect
the fluctuation spectrum. This unavoidable interface motion reflects
the fact that the semi-grand canonical Monte Carlo method models an
open system with neutral equilibrium between the phases. In this work
this problem was addressed by imposing an additional constraint maintaining
a fixed average composition of the two-phase system.

The simulation block had the approximate dimensions 28.6 by 50.0 by
1.4 nm with all-periodic boundary conditions and contained about 180,000
atoms. The $z$-dimension was relatively small to realize the ribbon-like
geometry of the interface. The block initially contained two plane
coherent (100) Ni/Ni$_{3}$Al interfaces normal to the $y$-axis.
Atomic interactions were modeled by the embedded-atom potential developed
in \citep{Mishin04a}. Monte Carlo simulations were run to equilibrate
the system at a chosen temperature. The trial moves of the Monte Carlo
process included (1) displacement of a randomly chosen atoms by a
random amount in a random direction and (2) reassignment of its chemical
species at random to either Ni or Al. The trial move was accepted
or rejected according to the Metropolis algorithm, namely, with the
probability $\exp(-\Phi/k_{B}T)$ if $\Phi>0$ and unconditionally
if $\Phi\leq0$. Here 
\begin{equation}
\Phi\equiv\Delta E\pm\Delta\mu\pm\dfrac{3}{2}k_{B}T\ln\dfrac{m_{\mathrm{Ni}}}{m_{\mathrm{Al}}},\label{eq:2}
\end{equation}
where $m_{\mathrm{Ni}}$ and $m_{\mathrm{Al}}$ are the atomic masses
\citep{Brown03a,Brown05}. The positive sign is used when Ni is replaced
by Al and the negative when Al is replaced by Ni. The logarithmic
term comes from integration of the probability distribution over the
momenta of all atoms, producing a pre-exponential factor proportional
to the product of all atomic masses to the power of 3/2. In the probability
ratio of two configurations, all masses cancel out except for the
atom whose species changes, producing a pre-exponential factor of
either $(m_{\mathrm{Ni}}/m_{\mathrm{Al}})^{3/2}$ or $(m_{\mathrm{Al}}/m_{\mathrm{Ni}})^{3/2}$.
Simultaneously with this Monte Carlo process, the zero-stress condition
was maintained in each Cartesian direction by a Rahman-Parrinello
algorithm. 

To constrain the average fraction of Al atoms in the system, $\bar{c}$,
to a preset target value $\bar{c}_{0}$, a feedback loop was created
between the current value of $\bar{c}$ and the imposed chemical potential
difference $\Delta\mu$. Namely, at each Monte Carlo step $n$, $\Delta\mu_{n}$
was modified by
\begin{equation}
\Delta\mu_{n}=\Delta\mu_{n-1}-\alpha\left(\dfrac{\bar{c}_{n-1}+\bar{c}_{n-2}}{2}-\bar{c}_{0}\right),\label{eq:4}
\end{equation}
where $\alpha$ is an adjustable constant. (One Monte Carlo step is
defined as the number of trial moves equal to the number of atoms.)
After a short transient, the system reached a regime in which its
composition fluctuated around $\bar{c}_{0}$ accompanied by fluctuations
of $\Delta\mu$ around a value $\Delta\mu_{0}$ corresponding to two-phase
coexistence. At each temperature, $\bar{c}_{0}$ was chosen so that
to create an equilibrium structure with nearly equal amounts of the
phases. The described feedback loop is technically similar to the
variance-constrained ensemble recently proposed by Sadigh et al.~\citep{Sadigh2012},
although our implementation of Eq.~(\ref{eq:4}) in the parallel
Monte Carlo code was different.

Once phase equilibrium was reached, the Monte Carlo run was continued
for approximately $10^{6}$ steps to save a set of statistically independent
snapshots and accurately compute average values of fluctuating parameters.

\begin{figure}
\noindent \begin{centering}
\includegraphics[clip,scale=0.9]{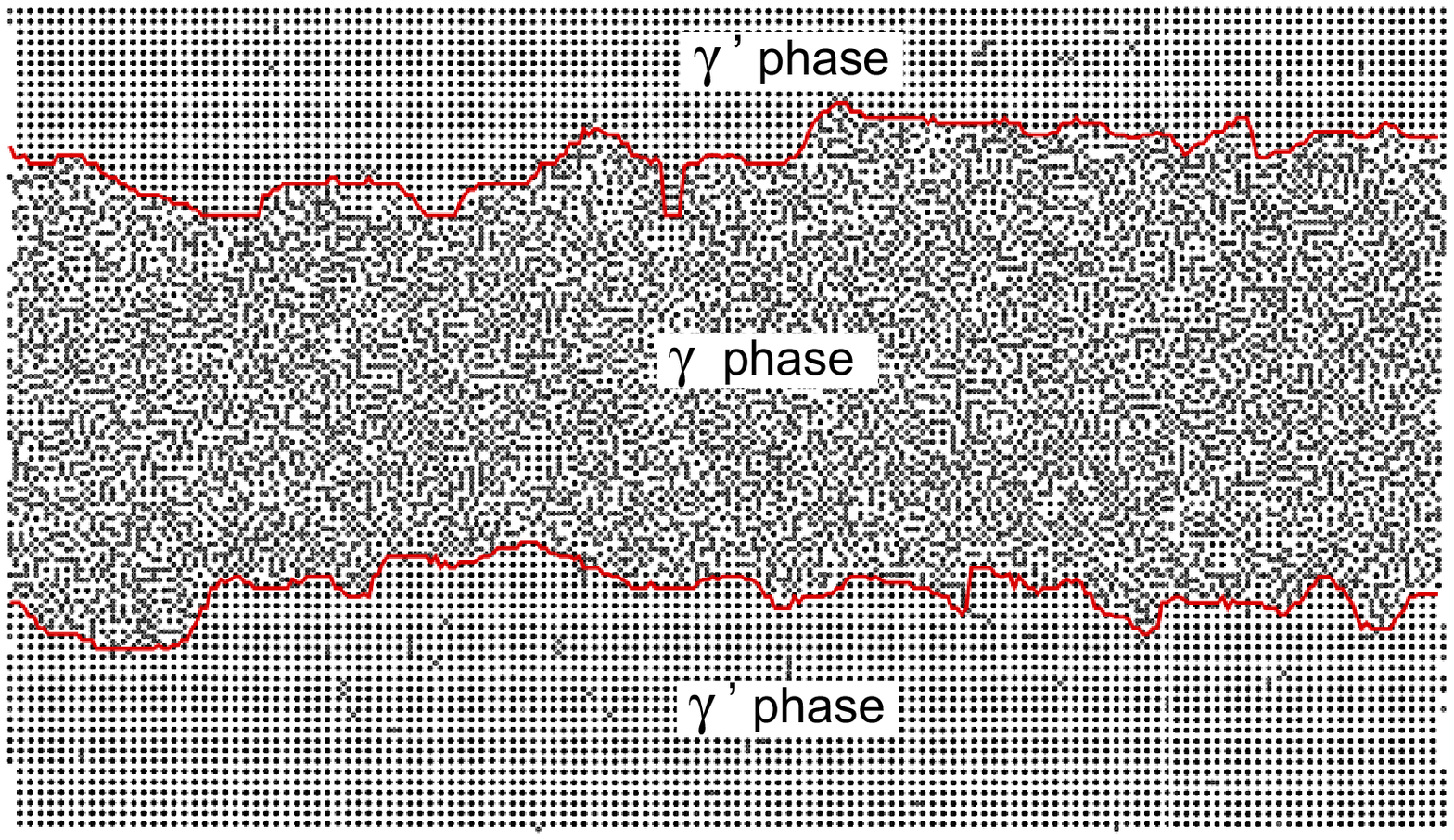}
\par\end{centering}

\caption{A typical snapshot of Monte Carlo simulations of $\gamma$ and $\gamma^{\prime}$
phases at the temperature of 700 K. The simulation block is projected
on the $x$-$y$ plane showing only Al atoms. The red lines show the
$\gamma/\gamma^{\prime}$ interfaces revealed by the visualization
method applied in this work.\label{fig:1}}
\end{figure}

At the post-processing stage, every snapshot was quenched to 0 K to
eliminate thermal noise. It was checked that the interfaces always
remained perfectly coherent. For each atom in a given snapshot, the
local chemical composition $\xi$ was computed as the fraction of
Al atoms in a sphere of a radius 4 $\textrm{\AA}$ centered at the
atom. The $\xi$ numbers were then averaged over bulk regions unaffected
by the interfaces to obtain the chemical compositions, $c_{\gamma}$
and $c_{\gamma^{\prime}}$, of the phases. To compute the interface
shape, the $x$-$y$ cross-section of the block was partitioned into
imaginary cells (pixels) of equal size using a 300$\times$400 mesh.
The $\xi$ numbers were averaged over each cell (including the $z$-direction)
to obtain a coarse-grained chemical composition $c$ assigned to each
pixel $(x_{i},y_{i})$. In each row of pixels parallel to the $y$-direction
(normal to the interface), $c$ was a function of the pixel ordinate
$y_{i}$. Namely, it remained nearly constant and equal to either
$c_{\gamma}$ or $c_{\gamma^{\prime}}$ inside the phases and changed
rapidly from one composition to the other  in the interface regions.
Using linear interpolation between neighboring pixels, for each interface
the coordinate $Y_{i}$ was found at which $c=(c_{\gamma}+c_{\gamma^{\prime}})/2$
and was identified with the interface position at the given $x_{i}$.
An example of interface profiles $(x_{i},Y_{i})$ is shown in Fig.~\ref{fig:1}
where the points are connected by straight segments for clarity. The
protrusions and zig-zags of the profile on the scale comparable to
the unit cell size depend on the visualization parameters and do not
represent in the physical shape of the interface. (The physical interface
width could be evaluated from composition profiles \emph{across} the
interface averaged over the snapshots, which was not pursued in this
work.) The discrete Fourier transformation was then applied to the
discrete function $Y_{i}(x_{i})$ and the Fourier amplitudes were
averaged over a few hundred snapshots to obtain $\left\langle a^{2}(k_{i})\right\rangle $
for a set of wavenumbers $k_{i}$. The phase compositions were also
averaged over the same set of snapshots.

We assumed that the term $\sigma^{\prime\prime}$ could be neglected
in comparison with $\sigma$. This cannot be proven rigorously but
is consistent with the very rough shape of the interface studied here.
The interface free energy was found by 
\begin{equation}
\sigma=\dfrac{k_{B}T}{A\left\langle a^{2}(k)\right\rangle k^{2}},\label{eq:3}
\end{equation}
where the right-hand slide is the slope of the plot $k_{B}T/A\left\langle a^{2}(k)\right\rangle $
versus $k^{2}$. 

\begin{figure}
\noindent \begin{centering}
\includegraphics[scale=0.59]{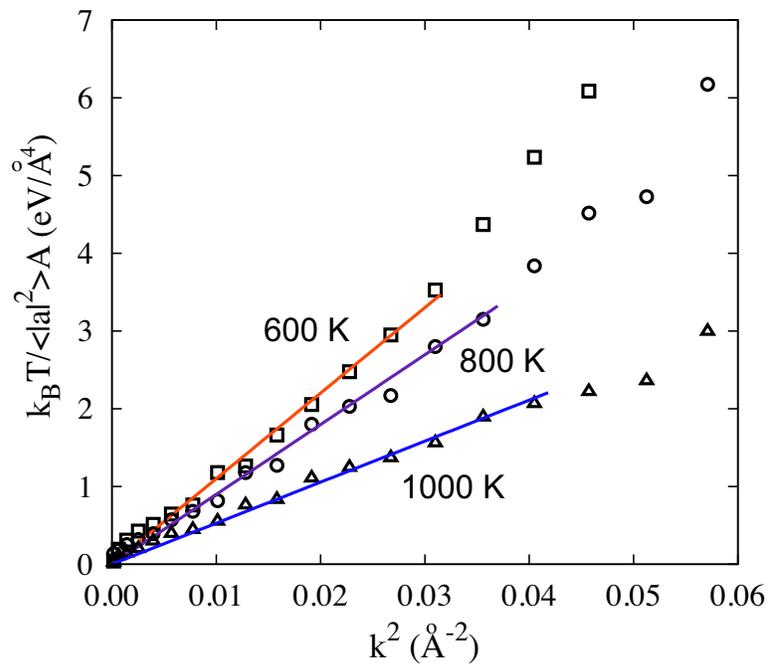}
\par\end{centering}

\caption{Typical plots of $k_{B}T/A\left\langle a^{2}(k)\right\rangle $ versus
$k^{2}$ at three temperatures. The straight lines are linear fits
in the long-wave range.\label{fig:2} }
\end{figure}

Fig.~\ref{fig:2} shows typical plots of $k_{B}T/A\left\langle a^{2}(k)\right\rangle $
versus $k^{2}$ and their mean-square linear fits. Significant deviations
from linearity are observed at large $k$ values when the amplitudes
become comparable to the pixel size and cannot be resolved accurately.
The linearity in the small $k$ region confirms that Eq.~(\ref{eq:3})
is a reasonable approximation. 

\begin{figure}
\noindent \begin{centering}
\includegraphics[scale=0.68]{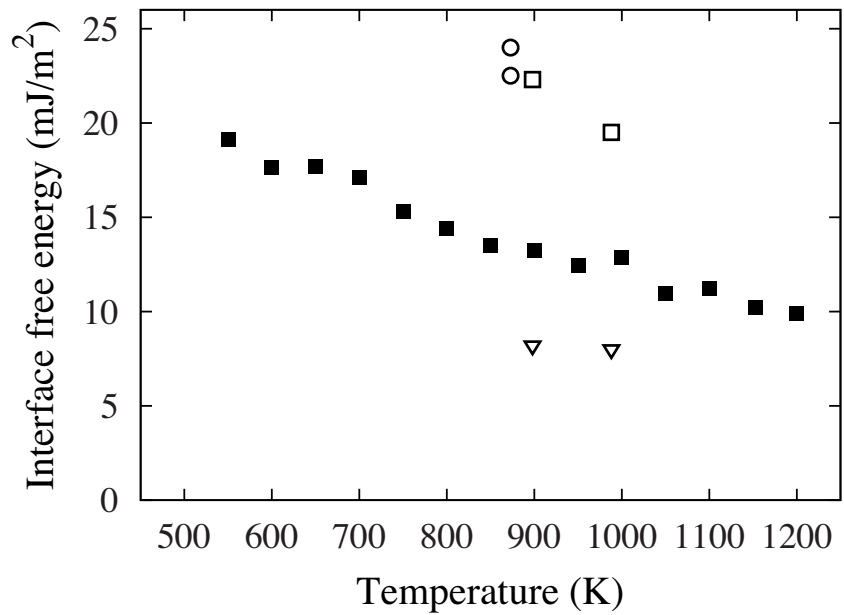}
\par\end{centering}

\caption{Comparison of (100) $\gamma/\gamma^{\prime}$ interface free energies
$\sigma$ computed by the capillary fluctuation method (filled squares)
in comparison with experimental data for an average interface orientation:
open triangles - Ni-Al system \citep{Ardell95}; open squares - Ni-Al
system with more accurate thermodynamics and TIDC model of coarsening
\citep{Ardell2011}; open circles - Ni-Al-Cr alloys with different
chemical compositions \citep{Booth-Morrison2008}.\label{fig:3}}
\end{figure}

The computed interface free energies were averaged over both interfaces
present in the system and are summarized in Fig.~\ref{fig:3}. The
plot starts at the temperature of 550 K because the amplitudes of
the capillary waves decrease with temperature and could not be resolved
below 550 K. On the high-temperature end, the simulations above 1200
K resulted in the rupture of the layers of the phases (Fig.~\ref{fig:1})
and formation of a single spherical $\gamma^{\prime}$ particle in
the $\gamma$ matrix. In the future, the temperature interval could
be extended to higher temperatures by using wider layers of the phases
to avoid their breakup.

Comparison of the calculated interface free energies with experiment
is not straightforward for a number of reasons. The experimental values
of $\sigma$ are back-calculated from observations of coarsening kinetics
of $\gamma^{\prime}$ particles. Such calculations rely on a kinetic
model and a thermodynamic description of the alloy. Different $\sigma$
values have been reported in the literature, depending on the choice
of the kinetic and thermodynamic models. This can partially explain
the significant scatter of the experimental data. For example, Fig.~\ref{fig:3}
includes earlier results \citep{Ardell95} based on the Lifshtiz-Slyozov-Wagner
(LSW) \citep{Lifshitz1961,Wagner1961} model of coarsening together
with more recent calculations \citep{Ardell2011} employing the trans-interface-diffusion-controlled
(TIDC) model \citep{Ardell05a} with more accurate thermodynamics,
both calculations utilizing the same set of experimental data. The
plot also includes $\sigma$ values for two ternary Ni-Al-Cr alloys
obtained by applying the Kuehmann and Voorhees model \citep{Kuehmann1996}
of coarsening to data measured by atom-probe tomography \citep{Booth-Morrison2008}.
Another factor to consider is the possible anisotropy of $\sigma$
suggested by previous calculations \citep{Mishin04a,Wang2009,Mao2012a}.
The experiments give $\sigma$ averaged over all possible orientations.
There is presently no experimental evidence of anisotropy of $\sigma$.
Small $\gamma^{\prime}$ particles are spherical and transform to
a cubic shape with (100) faces when they reach a certain size. However,
this shape transformation is caused by mistfit strains and the elastic
anisotropy of the material and cannot be interpreted as evidence that
$\sigma_{(100)}$ is smaller than for other crystallographic orientations
(see discussion in \citep{Ardell2011}). On the other hand, the magnitude
of $\sigma$ is so small that it might still be anisotropic without
affecting the particle shape. This possible anisotropy of $\sigma$
also makes the comparison of the computed $\sigma_{(100)}$ values
with average experimental values somewhat ambiguous.

Nevertheless, the computed $\sigma$ values are approximately in the
same ballpark as the known experimental data: 10 to 20 mJ/m$^{2}$.
Although the agreement is imperfect, it is non-trivial given the extremely
small magnitude of $\sigma$. Indeed, the computed $\sigma$ is about
1-2\% of the Al surface energy, below 1\% of the Ni surface energy,
and less than 1/3 of the surface tension of water at room temperature.

\begin{figure}
\noindent \begin{centering}
\includegraphics[scale=0.55]{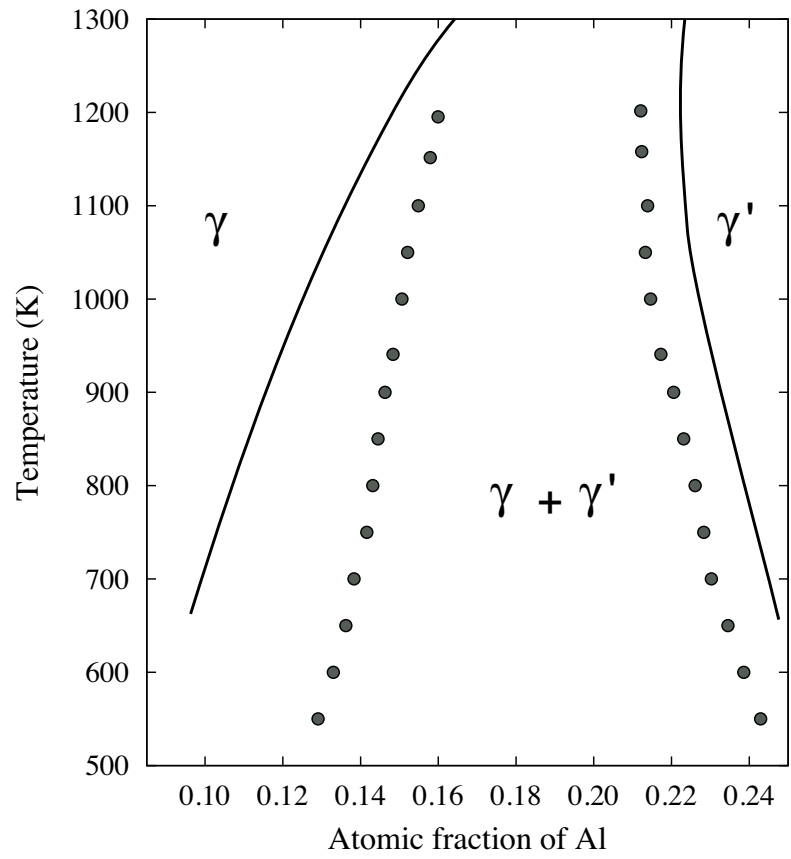}
\par\end{centering}

\caption{$\gamma$ and $\gamma^{\prime}$ solvus lines on the Ni-Al phase diagram
calculated in this work (points) in comparison with experiment \citep{Li1997,Ardell2000,Ma2003}
(lines).\label{fig:4} }

\end{figure}

The main source of error in the reported $\sigma$ is probably rooted
in the inaccuracy of the atomistic potential \citep{Mishin04a}. As
an illustration, Fig.~\ref{fig:4} displays the computed $\gamma$
and $\gamma^{\prime}$ solvus lines on the Ni-Al phase diagram. The
phase compositions $c_{\gamma}$ and $c_{\gamma^{\prime}}$ at different
temperatures were obtained as a side product in the calculations of
the interface profiles. Both solvus lines agree with the previous
calculation \citep{Mishin04a} by a different method using the same
potential, but show significant deviations from the experimental phase
diagram \citep{Li1997,Ardell2000,Ma2003}. In particular, the compositional
width of the $(\gamma+\gamma^{\prime})$ field on the phase diagram
is more narrow than in experiment, suggesting that the computed $\sigma$
values are likely to be slightly underestimated. The accuracy of the
calculations could be improved by developing a potential capable of
reproducing the solvus lines in better agreement with experiment.
Another factor neglected in this work is the effect of coherency strains.
At several temperatures, the calculations were repeated by fixing
the $x$-dimension of the simulation block at values creating slight
lateral tensions or compressions on the level of $\pm0.2$\%, modifying
the coherency strains in the phases. No effect on $\sigma$ was detected
within the statistical errors of the calculations. However, if the
accuracy of the method can be increases in the future, a way should
also be found to take into account the coherency strains. 

In conclusion, this work demonstrates that the capillary fluctuation
method in conjunction with Monte Carlo simulations has a potential
as a means of predicting free energies of $\gamma/\gamma^{\prime}$
interfaces and possibly other solid-solid interfaces with low free
energies. The small magnitude of $\sigma$ which makes it difficult
to compute it by thermodynamic integration or similar methods leads
to relatively large amplitudes of capillary waves and makes them amenable
to spectral analysis allowing the extraction of $\sigma$. The values
of $\sigma$ computed in this work are in reasonable agreement with
existing experimental data and show a significant decrease with temperature,
approaching \textasciitilde{}10 mJ/m$^{2}$ at high temperatures (Fig.~\ref{fig:3}).

I would like to thank A.~J.~Ardell, T.~Frolov, A.~Karma, D.~N.~Seidman
and V.~I.~Yamakov for reading the manuscript and providing helpful
comments. I am especially grateful to V.~I.~Yamakov for developing
the parallel Monte Carlo code and making it available for this work.
This research was sponsored by the National Institute of Standards
and Technology, Material Measurement Laboratory, the Materials Science
and Engineering Division. 


\newpage{}\clearpage{}

\end{document}